\documentclass{article}
\usepackage{spconf,amsmath,graphicx}
\usepackage{times}
\usepackage{epsfig}
\usepackage{subcaption, bm}
\usepackage{amssymb}
\usepackage{array}
\usepackage[table]{xcolor}
\usepackage{multirow}
\usepackage{url}

\graphicspath{{images/}}

\DeclareMathOperator*{\argmin}{arg\,min}

\newcolumntype{K}[1]{>{\centering\arraybackslash}p{#1}}

\title{Clonability of anti-counterfeiting printable graphical codes: \\ a machine learning approach}
\name{O. Taran, S. Bonev and S. Voloshynovskiy \thanks{S. Voloshynovskiy is a corresponding author. The research was supported by the SNF project No. 200021\_182063.}}
\address{University of Geneva, Department of Computer Science, Stochastic Information Processing Group \\
7, route de Drize, 1227 Geneva, Switzerland}
\begin{document}
%
\maketitle
\begin{abstract}
%
In recent years, printable graphical codes have attracted a lot of attention enabling a link between the physical and digital worlds, which is of great interest for the IoT and brand protection applications. The security of printable codes in terms of their reproducibility by unauthorized parties or clonability is largely unexplored. In this paper, we try to investigate the clonability of printable graphical codes from a machine learning perspective. The proposed framework is based on a simple system composed of fully connected neural network layers. The results obtained on real codes printed by several printers demonstrate a possibility to accurately estimate digital codes from their printed counterparts in certain cases. This provides a new insight on scenarios, where printable graphical codes can be accurately cloned.
%
\end{abstract}
\section{Introduction}
\label{sec:intro}

Counterfeiting of physical objects is a very important problem for the modern economies. 
There exist several techniques to protect original products against falsification and to provide a link between a physical object and its digital representation in centralized or distributed databases. This link can be implemented via \textit{overt channels}, like personalized codes reproduced on products either directly or in a form of coded symbologies like 1D and 2D codes or \textit{covert channels}, like invisible digital watermarks embedded in images or text or printed by special invisible inks. However, it is crucial to provide a non-clonability of this link to avoid any false acceptance of fake objects as authentic ones. Two most well known technologies that claim to ensure such a non-clonability are Physical Unclonable Functions (\textit{PUFs}) \cite{Voloshynovskiy:2008:SECSI} and Printable Graphic Codes (\textit{PGC}) that originate from the work \cite{picard2004digital}. The theoretical comparison of \textit{PUFs} and \textit{PGC} is given in \cite{Slava_ICASSP16_PUF}. In this paper we focus on the clonability of \textit{PGC}. 
The deployment of \textit{PGC} has a lot of advantages and attracts many industrial players and governmental organizations. Nevertheless, the claimed non-clonability of \textit{PGC} remains largely unexplored besides some rare exceptions \cite{baras2013towards, phan2013document}.

The main goal of this paper is to investigate the resistance of \textit{PGC} to clonability attacks. The overwhelming majority of such attacks can be split into two main groups: \textbf{(a)} \textit{hand-crafted} attacks, which are based on the experience and know-how of the attackers and \textbf{(b)} \textit{machine learning} based attacks, which use training data to create clones of the original codes. 

In this paper, we focus on the investigation of \textit{machine learning} based attacks due to the recent advent in the theory and practice of machine learning tools. Growing popularity and remarkable results of deep neural network (DNN) architectures in computer vision applications motivated us to investigate the clonability of \textit{PGC} using these architectures trained for different classes of printers. In our study, we assume that the detection mechanism of defender is also unknown, thus making our attack universal in this sense. 

\begin{figure}[t!]
\begin{center}
\includegraphics[width=1\linewidth]{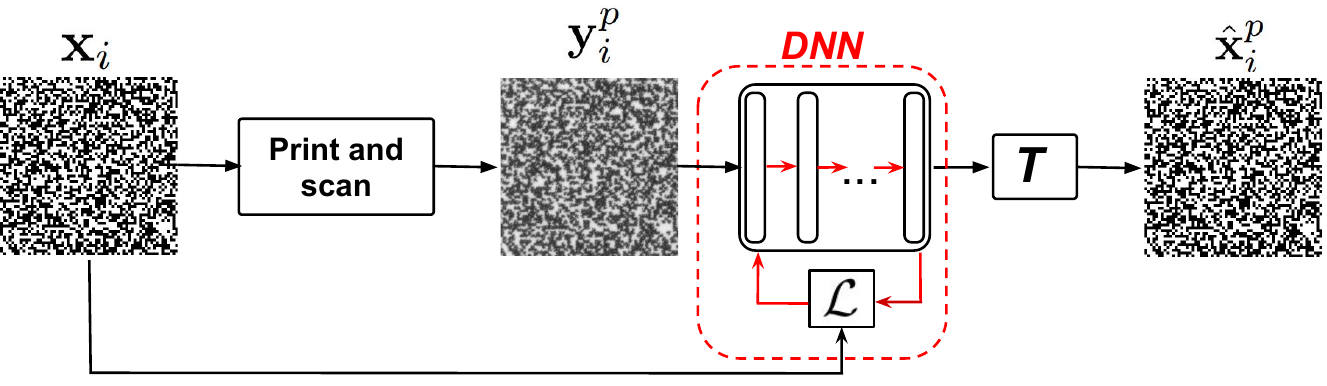}
\end{center}
\caption{Training procedure based on training samples (${\bf x}_i, {\bf y}_i^p$).}
\label{fig:training_procedure}
\vspace{-0.5cm}  
\end{figure}
Therefore, the main contributions of this paper are:

\indent - we investigate the clonability of printable graphical codes using machine learning based attacks;

\indent - we examine the proposed framework on real printed codes reproduced with 4 printers;

\indent - we empirically demonstrate a possibility to sufficiently accurately clone the \textit{PGC} from their printed counterparts in certain cases.
\begin{table*}[th!]
\renewcommand*{\arraystretch}{1.2}
\begin{center}
\begin{tabular}{c|cK{3.2cm}K{3.2cm}K{3.2cm}K{3.2cm}}
& Printer & Scanned original & Original & Reconstructed (\textit{BN}) & Difference \\ \hline
\multirow{ 6}{*}{\rotatebox{90}{Laser printers}} & & & & & \\[-0.4cm]
& SA & 
	\begin{minipage}{.2\textwidth}
    \centering
      \includegraphics[width=0.75\linewidth]{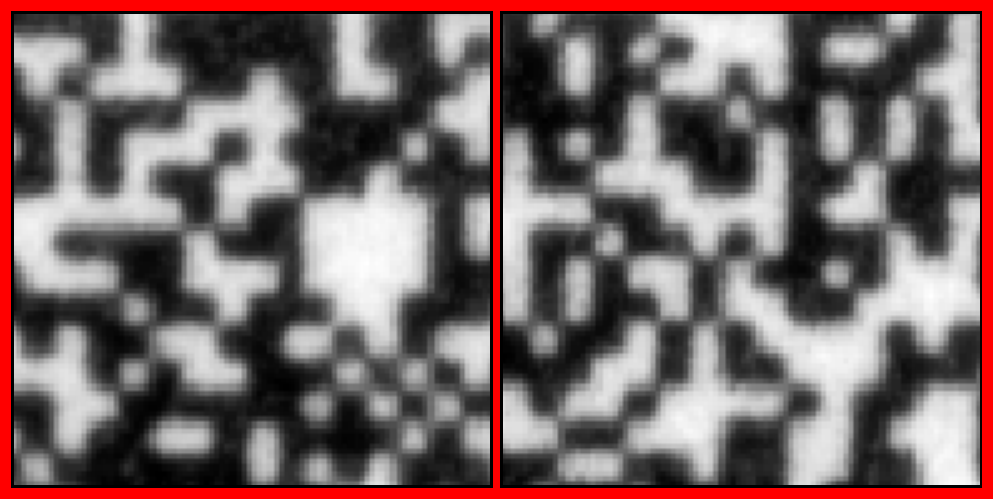}
    \end{minipage}  
    & 
	\begin{minipage}{.2\textwidth}
      \centering	
      \includegraphics[width=0.75\linewidth]{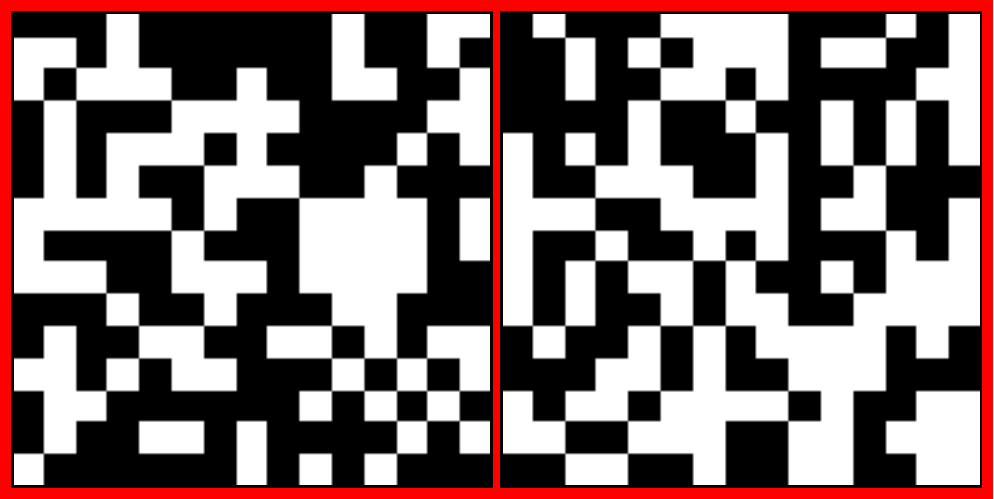}
    \end{minipage}      
    &  
	\begin{minipage}{.2\textwidth}
      \centering
      \includegraphics[width=0.75\linewidth]{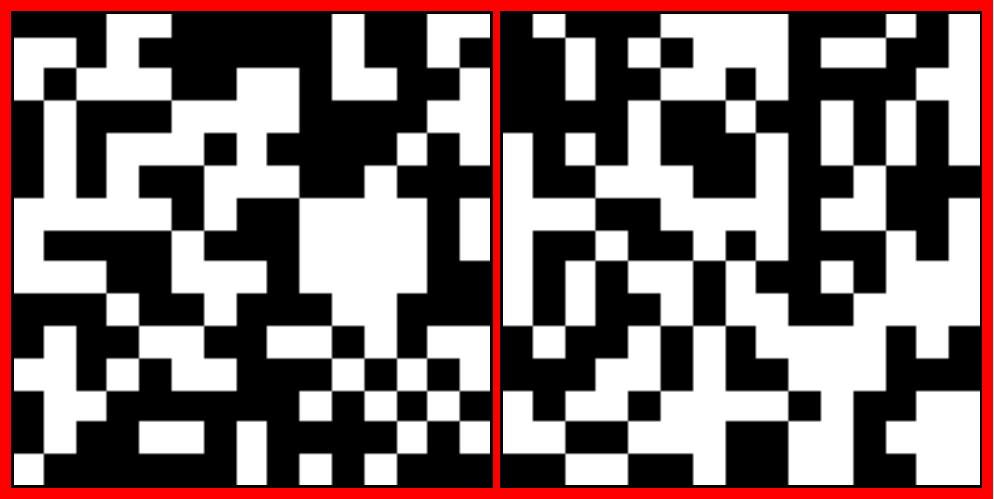}
    \end{minipage}      
    & 
	\begin{minipage}{.2\textwidth}
      \centering
      \includegraphics[width=0.75\linewidth]{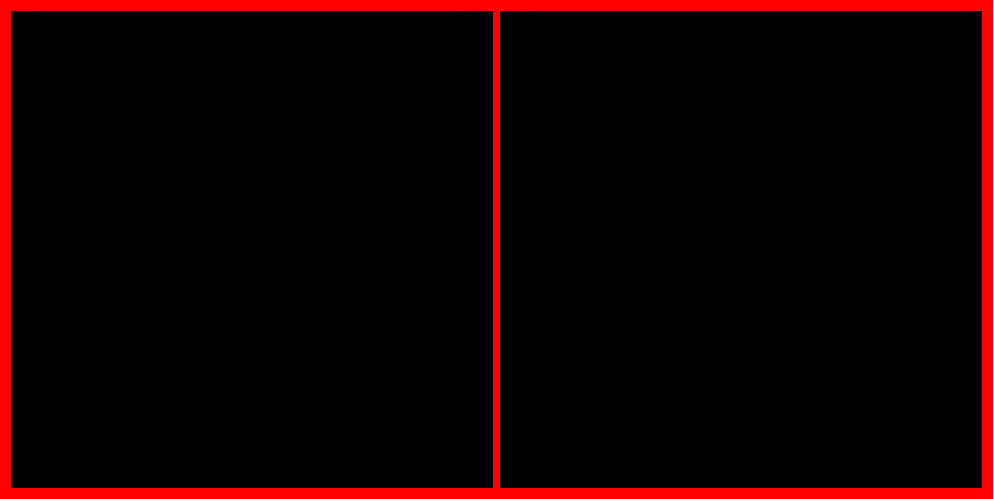}
    \end{minipage}      
    \\
& & & & & \\[-0.45cm] 
& & & & & \\[-0.45cm]
& LX & 
	\begin{minipage}{.2\textwidth}
    \centering
      \includegraphics[width=0.75\linewidth]{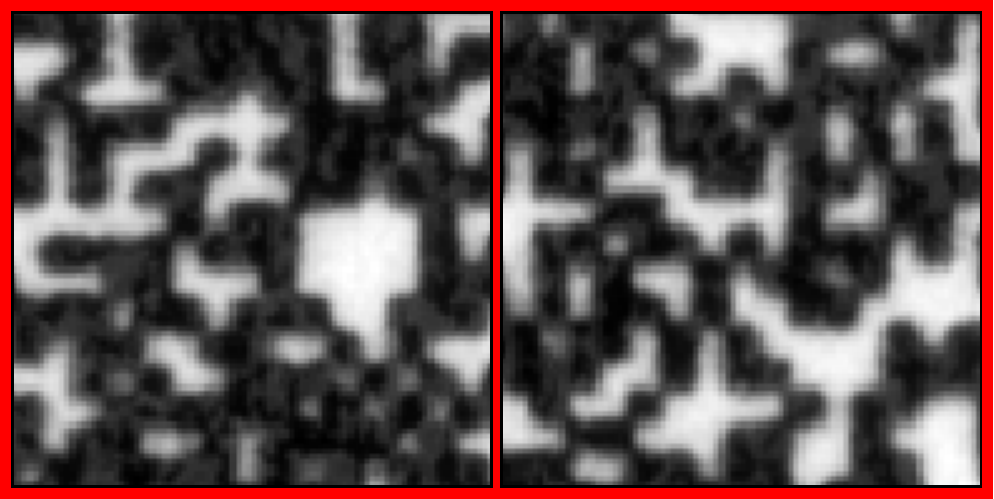}
    \end{minipage}  
    & 
	\begin{minipage}{.2\textwidth}
      \centering	
      \includegraphics[width=0.75\linewidth]{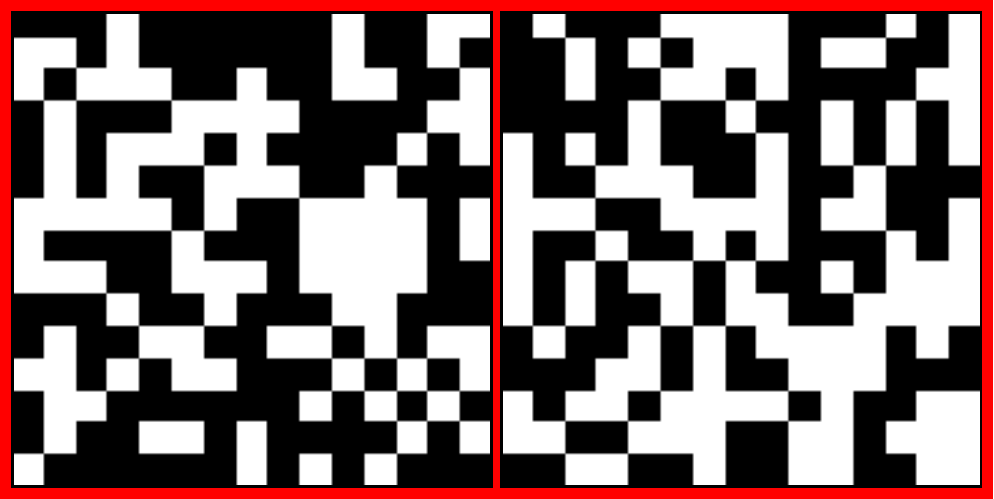}
    \end{minipage}      
    &  
	\begin{minipage}{.2\textwidth}
      \centering
      \includegraphics[width=0.75\linewidth]{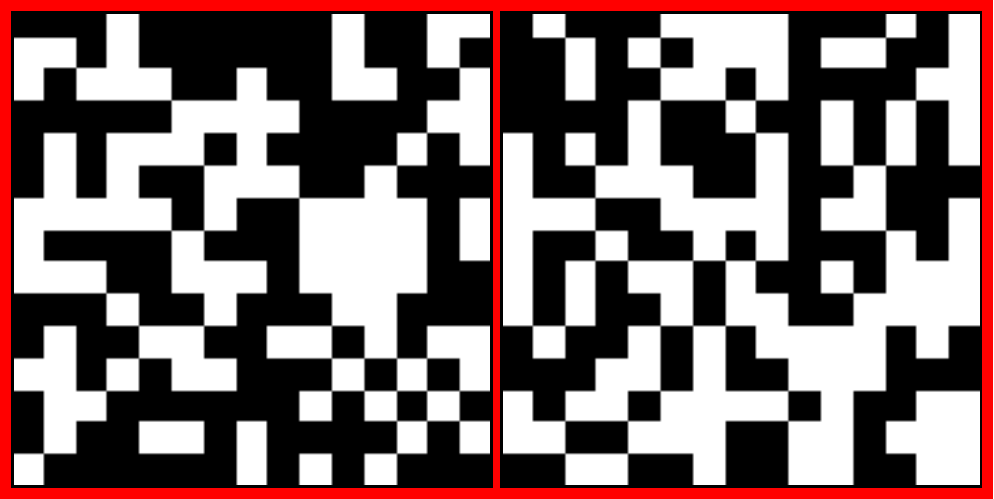}
    \end{minipage}      
    & 
	\begin{minipage}{.2\textwidth}
      \centering
      \includegraphics[width=0.75\linewidth]{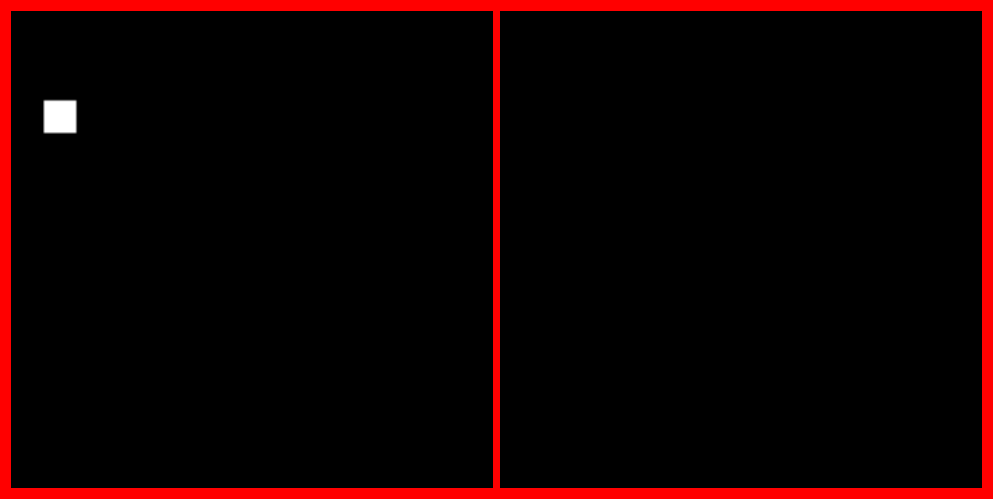}
    \end{minipage}      
    \\
& & & & & \\[-0.4cm] 
\hline
\multirow{6}{*}{\rotatebox{90}{Inkjet printers}} & & & & & \\[-0.4cm]
& HP & 
	\begin{minipage}{.2\textwidth}
    \centering
      \includegraphics[width=0.75\linewidth]{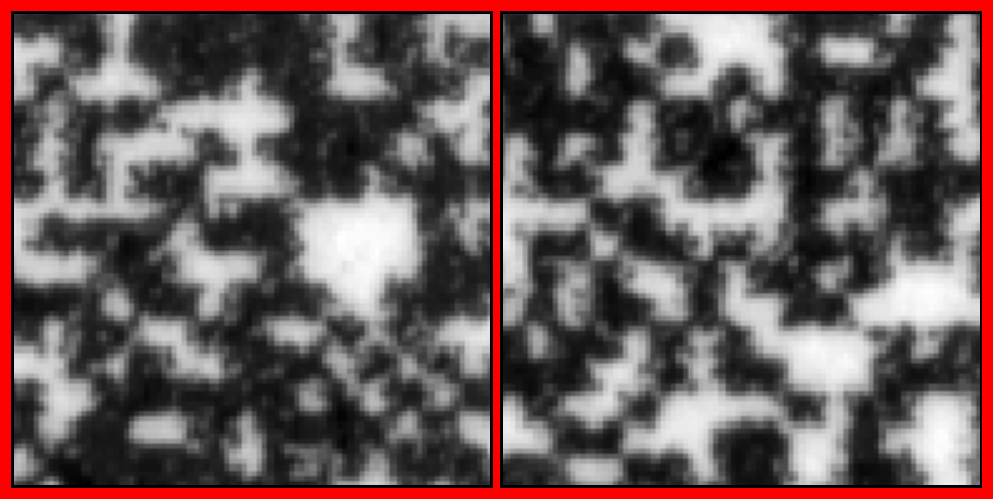}
    \end{minipage}  
    & 
	\begin{minipage}{.2\textwidth}
      \centering	
      \includegraphics[width=0.75\linewidth]{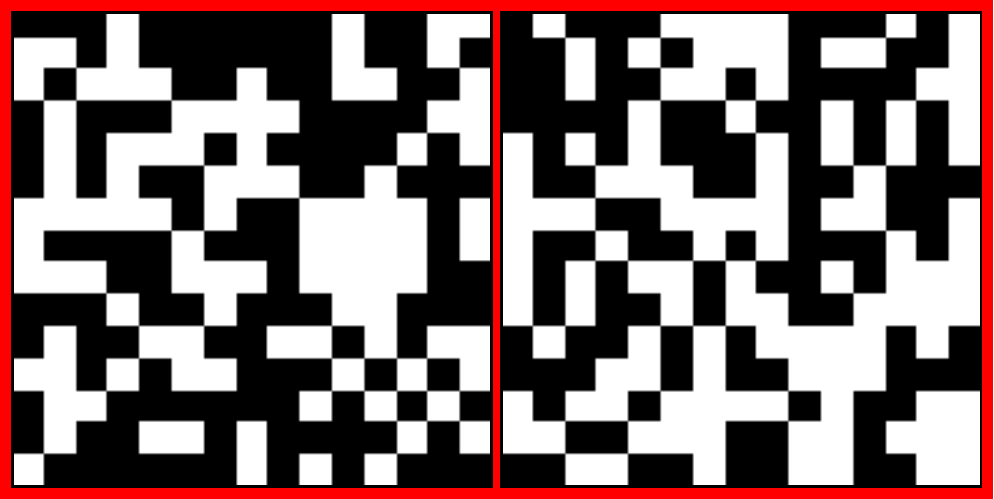}
    \end{minipage}      
    &  
	\begin{minipage}{.2\textwidth}
      \centering
      \includegraphics[width=0.75\linewidth]{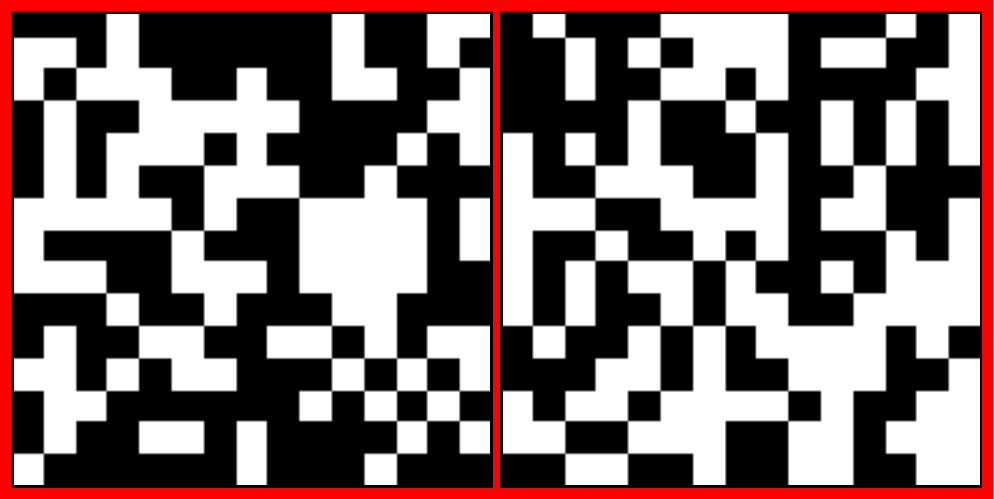}
    \end{minipage}      
    & 
	\begin{minipage}{.2\textwidth}
      \centering
      \includegraphics[width=0.75\linewidth]{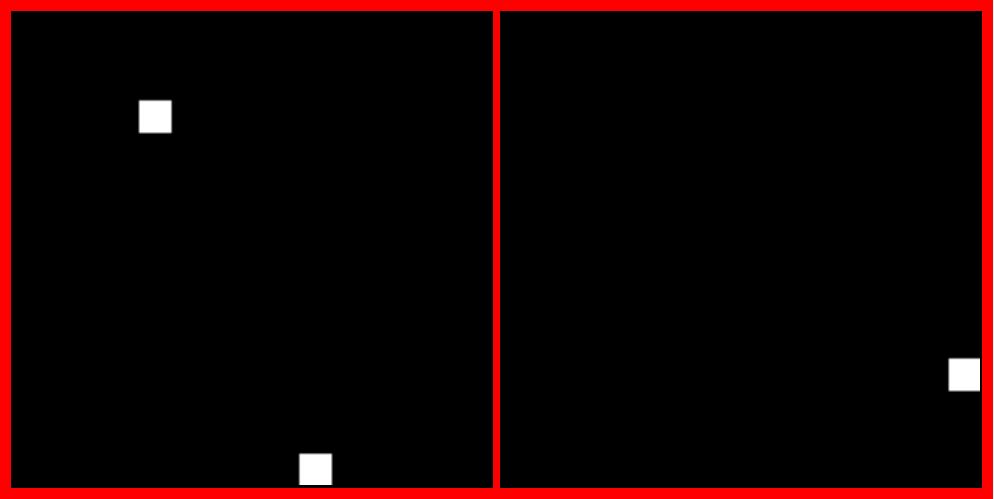}
    \end{minipage}      
    \\
& & & & & \\[-0.45cm] 
& & & & & \\[-0.45cm]
& CA & 
	\begin{minipage}{.2\textwidth}
    \centering
      \includegraphics[width=0.75\linewidth]{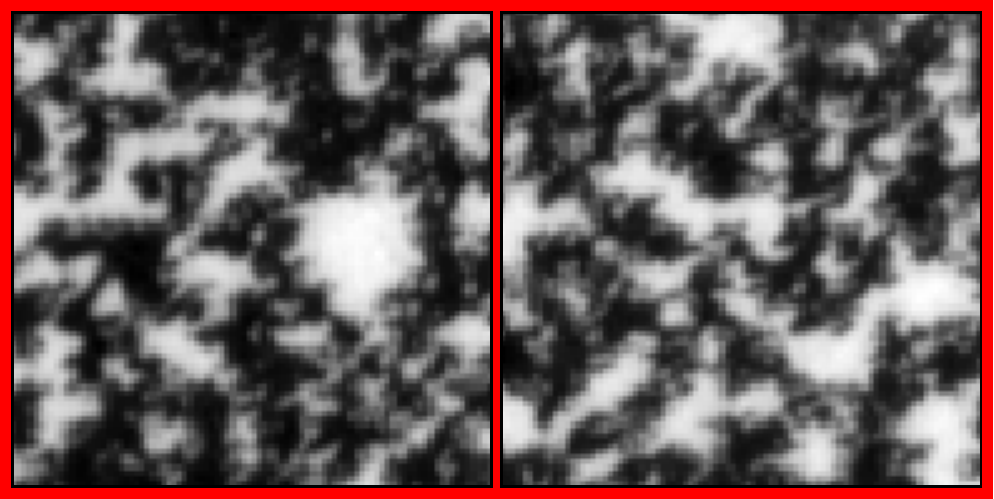}
    \end{minipage}  
    & 
	\begin{minipage}{.2\textwidth}
      \centering	
      \includegraphics[width=0.75\linewidth]{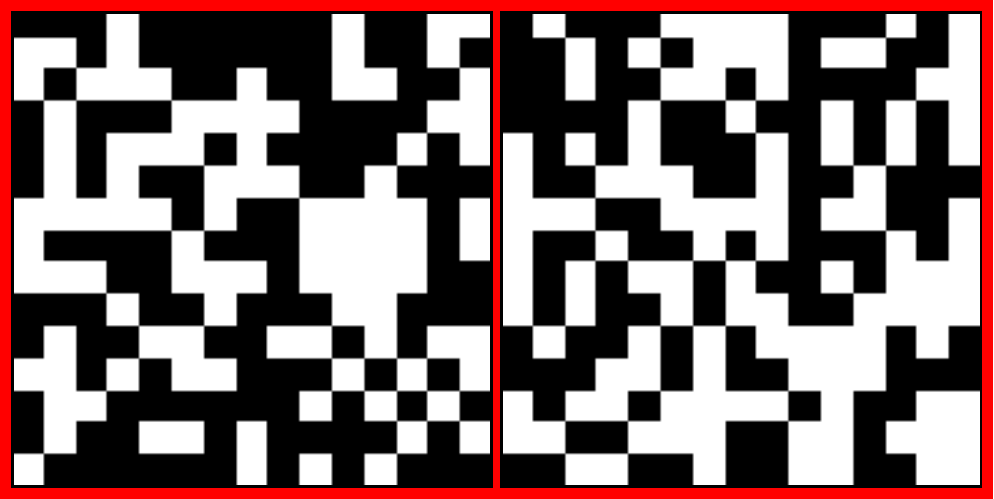}
    \end{minipage}      
    &  
	\begin{minipage}{.2\textwidth}
      \centering
      \includegraphics[width=0.75\linewidth]{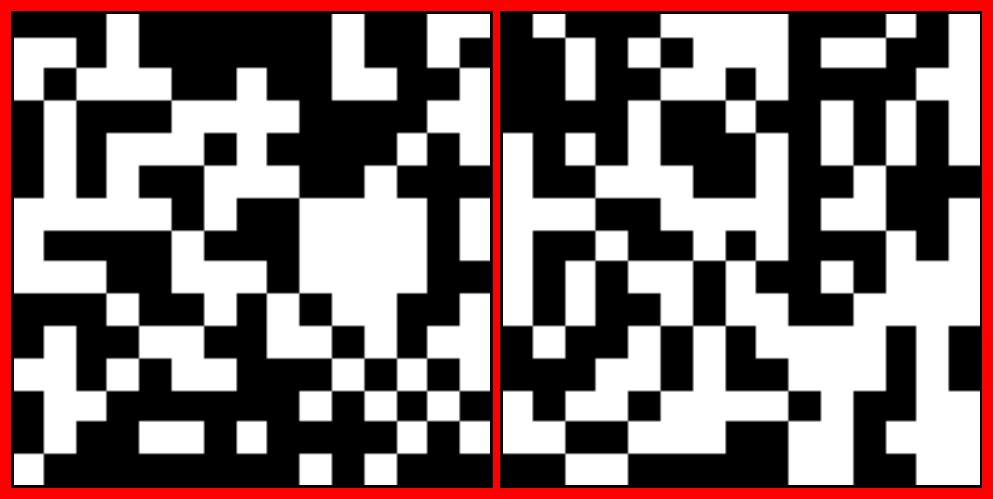}
    \end{minipage}      
    & 
	\begin{minipage}{.2\textwidth}
      \centering
      \includegraphics[width=0.75\linewidth]{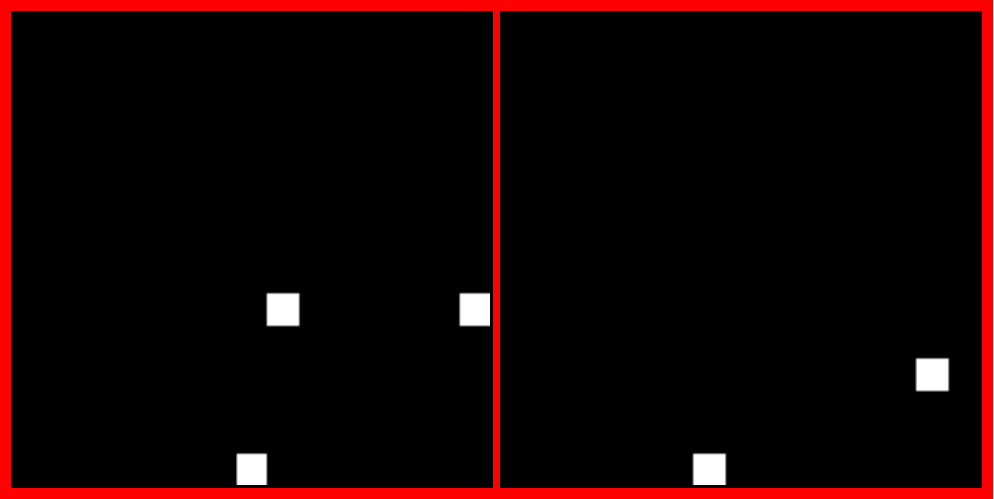}
    \end{minipage}    
    \\[0.5cm] \hline
\end{tabular}
\caption{Examples of attacks against PGC: two samples of scanned codes, the estimates produced by \textit{BN} model and the difference between the original and estimated codes.}
\label{tab:ae_rsults}
\end{center}
\vspace{-0.75cm}  
\end{table*}

The reminder of this paper is organized as follows: Section \ref{sec:problem statement} introduces a problem formulation and gives the details about the used test patterns, DNN architecture, training and test processes. Section \ref{sec:Experiments and Discussion} provides the details about the used printers, scanner and used dataset. The obtained empirical results and their analysis complete Section \ref{sec:Experiments and Discussion}. Section \ref{sec:conclusion} concludes the paper. 
\section{Problem formulation}
\label{sec:problem statement}
%
%
%
In our set up, we assume that the training dataset of the pairs of original (digital) ${\bf X} = \{{\bf x}_1, ..., {\bf x}_M\} \in \{ 0, 1\}^{N \times M}$ and printed codes ${\bf Y}^p = \{{\bf y}_1^p, ..., {\bf y}_M^p\} \in \mathbb{R}^{N \times M}$ reproduced with popular high resolution printing technologies denoted as $p=\{1,..., P\}$ is given. We assume ${\bf y}_i^p \in \mathbb{R}^N$ due to the scanning of printed codes in general, or ${\bf y}_i^p \in \{0, ..., 255\}^N$ in particular. The goal of the attacker is to obtain an accurate estimation of the original digital codes $\hat{{\bf X}}^p = \{\hat{{\bf x}}_1^p, ..., \hat{{\bf x}}_M^p\} \in \{ 0, 1\}^{N \times M}$ for each printing technology $p$. Mathematically, it can be formulated as an estimate:
\begin{equation}
\hat{{\bf x}}_i^p = f_{\bm{\vartheta}^p}({\bf y}_i^p),
\label{eq:1}
\end{equation}
\noindent where $i=1, ..., M$ and, in general case, $f_{\bm{\vartheta}^p}(.)$ can be any trainable function with the parameters $\bm{\vartheta}^p$. 
\subsection{Deep neural network models}
\label{subsec:DNN}
Nowadays, DNN technologies offer ample opportunities for training $f_{\bm{\vartheta}^p}(.)$ in a form of parametrized deep architectures. We investigate the possibilities of two types of the DNN architectures for solving problem (\ref{eq:1}).  The first one is based on several fully connected layers of the same size as the input data. Inspired by the fundamental role of autoencoders \cite{hinton2006reducing, goodfellow2014generative} in unsupervised learning, we investigate a system based on a "bottleneck" structure, where the dimensionality of the layers is reduced to the middle layer and then expanded to the full dimensionality in the output layer.

For both architectures the general schema of the training procedure is shown in Fig. \ref{fig:training_procedure}. For the given pairs of original and printed codes $({\bf X}, {\bf Y}^p)$, it can be formulated as: 
\begin{equation}
\hat{\bm{\theta}}^p = \argmin_{\bm{\theta}^p} \sum_{i=1}^M { \mathcal{L}\big({\bf x}_i, \phi_{\bm{\theta}^p}({\bf y}^p_i)\big)} + \lambda \Omega_{\bm{\theta}^p}(\bm{\theta}^p),
\label{eq:2}
\end{equation}
\noindent where $\mathcal{L}(.)$ is a loss function, $\phi_{\bm{\theta}^p}$ is a trained model, $\bm{\theta}^p$ represents the parameters of the trained model for chosen printer $p$ and $\Omega_{\bm{\theta}^p}(.)$ is a regularizer for the model parameters. In the case of an "bottleneck" model  $\phi_{\bm{\theta}^p} = \phi_{\bm{\theta}^p_D}(\phi_{\bm{\theta}^p_E}(.))$, $\bm{\theta}^p = (\bm{\theta}^p_E, \bm{\theta}^p_D)$ with $\bm{\theta}^p_E$ and $\bm{\theta}^p_D$ denoting the parameters of encoder and decoder parts, respectively. 

In the vast majority of cases the original digital codes are binary. However, training the DNN model with binary output is not a trivial task due to the difficulties with derivatives and vanishing of the gradients. For this reason in our framework the output of the trained models is real. The binarization of the regenerated codes is performed via a simple thresholding with an optimal threshold estimated on the validation subset. Therefore, the function $f_{\bm{\vartheta}^p}$ in the equation (\ref{eq:1}) is: 
\begin{equation}
f_{\bm{\vartheta}^p}(.) = T_{t^p}(\phi_{\bm{\theta}^p}(.)),
\label{eq:3}
\end{equation}
\noindent where $T(.)$ is a thresholding function with the threshold parameter $t^p$ and $\bm{\vartheta}^p = (\bm{\theta}^p, t^p)$.

At the test phase, the scanned sample ${\bf y}^p_i$ is passed through the pre-trained DNN model. The estimation of the original code $\hat{\bf x}_i^p$ is obtained after thresholding $T$ of the DNN output. The estimated code is printed and scanned on the corresponding equipment and the final decision about the code authenticity is made based on a chosen similarity measure $d(.)$ between the original and printed estimated codes. 

\subsection{Test pattern}
\label{subsec:test_patterns}

We used the \textit{DataMatrix} symbology consisting of 72x72 modules, which is described in the international standard ISO/IEC 16022 \cite{ISO2006}. To obtain a random bit distribution, the finder patterns were removed and only the mapping matrix of 64x64 modules was used for the printing tests.     



It should be pointed out that although the \textit{DataMatrix} code was initially proposed as an overt feature for personalisation applications, the chosen parameters of this code closely resemble those of the recently proposed \textit{PGC} that might  be equivalently printed up to a resolution of 2400 dpi. In our study we do not target to investigate the clonability of some particular \textit{PGC}, but rather to demonstrate a general approach applicable to the majority of \textit{PGC} designed with identical modulation principles.
%
\section{Experiments and Discussion}
\label{sec:Experiments and Discussion}
\textbf{Digital printers.} To evaluate the clonability aspects of PGC based on \textit{DataMatrix} modulation and to investigate the influence of the printing technologies we use 4 digital printers: 2 inkjet printers HP OfficeJet Pro 8210 (\textit{HP}) and Canon PIXMA iP7200 (\textit{CA}) and 2 laser printers Lexmark CS310 (\textit{LX}) and Samsung Xpress 430 (\textit{SA}). 

It should also be pointed out that up to our best knowledge there does not exist any accurate mathematical model describing the process of interaction between the ink and substrate (paper) besides some experimental studies as for example in \cite{Villan:TIFS2006, phan2013document}. Due to the fact that in all our experiments we use the same paper, we skip this parameter in our models for simplicity, but the impact of the substrate can be investigated in a similar manner to the proposed one. Therefore, without loss of generality we assume that the model of printing process is unknown and is not required for our attack strategy.

\textbf{DNN architectures.} In our experiments we use two types of DNN architectures with the same input size equals to 576: 

\textbf{1.} \textit{FC}: fully connected DNN with 2, 3 and 4 hidden layers (hereafter referred to as \textit{FC 2}, \textit{FC 3} and \textit{FC 4}). The size of each layer equals to the input size. 

\textbf{2.} \textit{BN}: "bottleneck" model with 2 fully connected hidden layers of size 256 and 128 at the encoder and decoder parts and a latent representation of size 36. 

In both cases we used $\ell_2$ norm as a loss function $\mathcal{L}(.)$ in (\ref{eq:2}).
%
%
The DNNs were implemented in Pytorch\footnote{https://github.com/taranO/clonability-of-printable-graphical-codes}. The training of the models was done on the \textit{Titan X} GPU card with the "Maxwell" architecture . All models were trained during 1 000 epochs with batch size equals to 128 and the learning rate equals $1\mathrm{e}{-3}$. 
\begin{table}[t!]
\renewcommand*{\arraystretch}{1.2}
\begin{center}
\begin{tabular}{ccccc}
\hline
\textbf{Method} & \textit{SA} & \textit{LX} & \textit{HP} & \textit{CA} \\ \hline 
\multicolumn{5}{c}{\textit{Pearson correlation}} \\ \hline
\textit{Thr}  & 0.774 & 0.766 & 0.742 & 0.704 \\ \hline 
\textit{FC} 2 & 0.995 & 0.994 & 0.982 & 0.981 \\ \hline 
\textit{FC} 3 & 0.994 & 0.994 & 0.982 & 0.983 \\ \hline 
\textit{FC} 4 & 0.994 & 0.995 & 0.981 & 0.982 \\ \hline 
\textit{BN}   & \textbf{0.996} & \textbf{0.996} & \textbf{0.986} & \textbf{0.984} \\ \hline 
\multicolumn{5}{c}{\textit{normalized Hamming distance}} \\ \hline
\textit{Thr} & 11 & 12 & 13 & 15 \\ \hline
\textit{FC} 2 & 0.22 & 0.24 & 0.93 & 0.98 \\ \hline 
\textit{FC} 3 & 0.23 & 0.24 & 0.90 & 0.85 \\ \hline 
\textit{FC} 4 & 0.24 & 0.23 & 0.95 & 0.90 \\ \hline
\textit{BN} & \textbf{0.21} & \textbf{0.22} & \textbf{0.69} & \textbf{0.76} \\ \hline
\end{tabular}
\caption{Regeneration accuracy with respect to original codes}
\label{tab:ae_reg_rsults}
\end{center}
\vspace{-0.9cm}
\end{table}
\begin{figure*}[t!]
	\centering
	\begin{subfigure}[t]{.245\textwidth}
      \centering
      \includegraphics[width=1.12\linewidth]{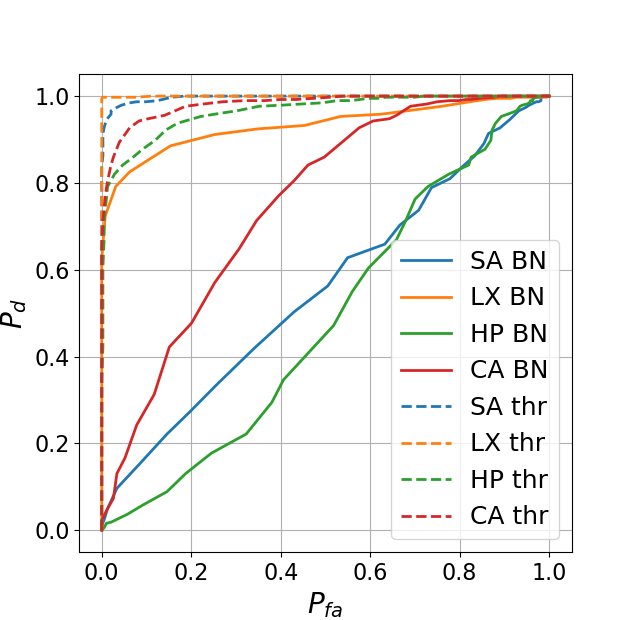}
      \caption{Pearson correlation}
      \label{fig:pearson_hist}
    \end{subfigure} 
    	\begin{subfigure}[t]{.245\textwidth}
      \centering
      \includegraphics[width=1.12\linewidth]{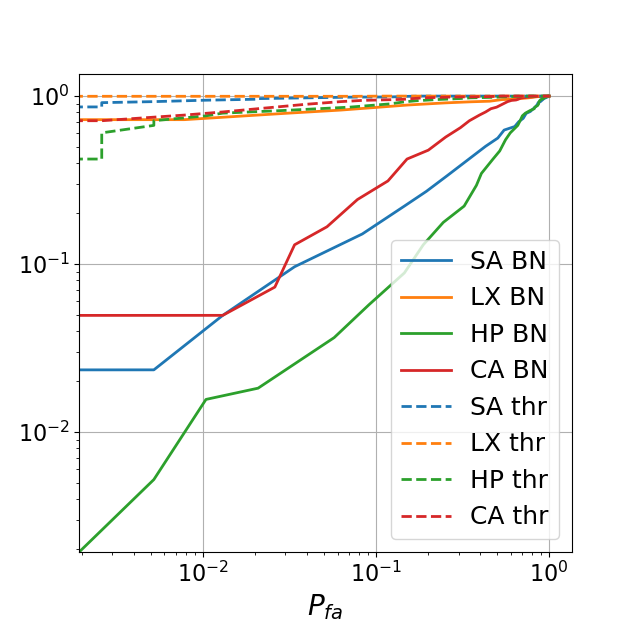}
      \caption{Pearson correlation in \textit{log} scale}
      \label{fig:pearson_hist}
    \end{subfigure} 
	\begin{subfigure}[t]{.245\textwidth}
      \centering
      \includegraphics[width=1.12\linewidth]{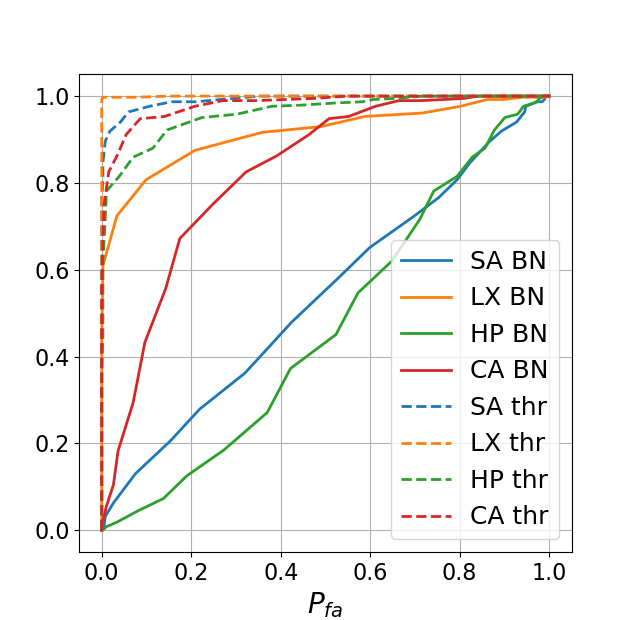}
      \caption{Hamming distance}
      \label{fig:xor_hist}
    \end{subfigure} 
	\begin{subfigure}[t]{.245\textwidth}
      \centering
      \includegraphics[width=1.12\linewidth]{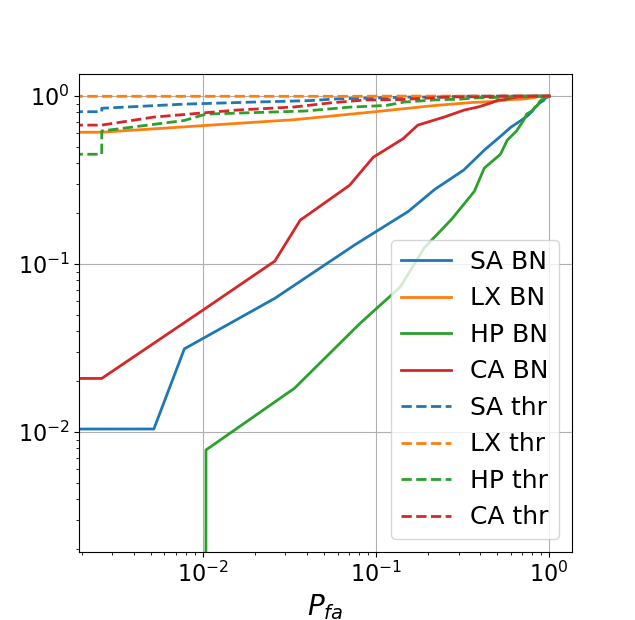}
      \caption{Hamming distance in \textit{log} scale}
      \label{fig:xor_hist}
    \end{subfigure} 
\caption{The ROC curves for \textit{Pearson correlation} and \textit{Hamming distance} between the original and fake printed codes estimated via \textit{\texttt{BN}} and \textit{\texttt{Thr}} methods. $P_d$ denotes to the probability of the correct detection and $P_{fa}$ is the probability of false acceptance.}
\label{fig:regeneration_roc}  
\vspace{-0.4cm}  
\end{figure*}

\textbf{PGC dataset.} The dataset of 384 original binary codes $\bf X$ of size $384 \times 384$ was generated according to the \textit{DataMatrix} standard described in Section \ref{subsec:test_patterns}. To obtain the dataset of the printed/scanned codes ${\bf Y}^p$ $p=\{1, ..., P\}$, the original codes were  printed on 4 printers ($P=4$) and scanned with the Epson Perfection V850 Pro scanner at 1200 ppi. The pairs $({\bf X}, {\bf Y}^p)$ were split into \textit{training} (100 images), \textit{validation} (50 images) and \textit{test} (234 images) subsets. Taking into account that the input size of the used DNN models equals 576, each image was split into non-overlapping blocks of size $24 \times 24$. Thus, the final size of the \textit{training} set is 25 600 sub-images, the \textit{validation} set contains 12 800 sub-images and the \textit{test} set consists of 59 904 sub-images.  

It should be pointed out that the training procedure is blind in the sense that we did not use any information about the principles of the \textit{DataMatrix} code generation. 

To evaluate the accuracy of the prediction of "regenerated" codes we use \textit{Pearson correlation} and normalized \textit{Hamming distance} between the original digital codes and the corresponding regenerated ones. The obtained results are presented in Table  \ref{tab:ae_reg_rsults}. Additionally to the DNN models, we perform the estimation from the printed codes via a simple thresholding (without DNN processing) similarly to \cite{baras2013towards, phan2013document, picard2004digital}. The obtained results correspond to the \textit{Thr} method in Table \ref{tab:ae_reg_rsults} and serve as baseline error. From the presented results, it is clear that the \textit{BN} architecture provides the best results. To provide more understanding how the codes look, we visualize the sub-blocks of size $84 \times 84$ from several codes for each printer and the estimations deploying the \textit{BN} as the best estimator in Table \ref{tab:ae_rsults}. 


To answer the question if the amount of errors in the \textit{BN} regenerated codes can be noticed by the defender and how the \textit{BN} results differ from the baseline estimation obtained via \textit{Thr} method, we printed our estimated codes for both cases on the same printers with the same parameters as the original codes and after that we scanned them on the same scanner. To evaluate the authenticity of the obtained results a number of metrics can be used. However, according to the authors in \cite{phan2013document} the most used one is a comparison of an original with a binarized or grey level version of the printed code. The authors in \cite{phan2013document} claim that the comparison with the grey level observations is preferable, since binarization is  a lossy transformation. In our evaluation we use \textit{Pearson correlation} between the originals and grey level printed codes. Additionally, we use normalized \textit{Hamming distance} to measure the accuracy of the logical symbol estimation in the originals and binarized printed codes. Using these statistics, we compute the ROC curves based on the probability of correct detection $P_d$ and the probability of false acceptance $P_{fa}$ via:
\begin{equation}
\begin{array}{lll}
P_d    & = & \textrm{Pr}\{ \alpha \cdot d(\bm{x}_i, \bm{y}_i^p) \ge \gamma | \mathcal{H}_0 \} \\
P_{fa} & = & \textrm{Pr}\{\alpha \cdot d(\bm{x}_i, \bm{y}_i^p) > \gamma | \mathcal{H}_1 \}, 
\end{array}
\label{eq:4}
\end{equation}
where $\gamma$ is the threshold, $d(.)$ is a similarity measure between the original and printed codes, $\mathcal{H}_0$ corresponds to the hypothesis that $\bm{y}_i^p$ is an authentic code and $\mathcal{H}_{1}$ is the hypothesis that $\bm{y}_i^p$ is a fake (cloned) code, $\alpha$ equals to 1 for the \textit{Pearson correlation} and to -1 for \textit{Hamming distance}. 

The obtained ROC curves are illustrated in Fig. \ref{fig:regeneration_roc}. It is easy to see that comparing with the baseline estimation via \textit{Thr} method, the system with \textit{BN} models makes the fake detection more difficult for defender. Particularly, as it can be noticed from Fig. \ref{fig:regeneration_roc}, in contrast to \textit{Thr} based estimation in the case of \textit{SA} and \textit{HP} printers, it is absolutely impossible to reliably distinguish the originals and fakes. For the \textit{SA} printer this result is evident due to the previously demonstrated high quality estimation. In the case of \textit{HP} printer such a result can be explained by the fact that, besides a quite big amount of errors in the estimated codes, the printing quality is relatively poor due to the high dot-gain. This leads to a sufficient amount of errors in the original printed codes that are masked in the printed fake codes due to the dot-gain effect. As a result, both codes become very close. In the case of the \textit{CA} printer, the ROC behaviour is superior, which is expected due to the previously demonstrated low quality estimation. The situation with the  \textit{LX} printer is the most interesting. From one point, the printing quality of this printer is a little bit worse than for the \textit{SA} printer and the obtained estimated error is not much higher. However, detailed analysis shows that the distributions of the errors between the codes is different. In the case of the \textit{SA} printer there are about 50\% of estimated codes without any mistake. This makes these fakes undistinguishable for the detector and the general quality of fake detection low. In the case of the \textit{LX} printer, the errors are distributed more or less uniformly between the codes, in the sense that almost each code has estimation errors. Due to the high printing quality this makes these codes "better" distinguishable for the detector. Nevertheless, it should be pointed out that the general level of false acceptance for the \textit{LX} printer is too high for practical use. For example, as can be seen from Fig. \ref{fig:regeneration_roc} for the probability of correct detection of around 0.95 the probability of false acceptance is 0.6. To have the $P_{fa}$ close to 0, one can achieve the $P_d$ of only about 0.6 - 0.8. For practical applications $P_d$ should not be less than 0.99 with the $P_{fa}$ not exceeding $10^{-6}$. From this we can conclude that the obtained results demonstrate the low resistance of the PGC based on DataMatrix modulation and similar codes to the \textit{machine learning} based clonability attacks.
%
\section{Conclusions}
\label{sec:conclusion}

In this paper we investigated the clonability of printable graphical codes using \textit{DataMatrix} modulation typical for many \textit{PGC} designs using \textit{machine learning} based attacks. We tested the proposed framework with two different DNN architectures on real printed data. We empirically proved the possibility to accurately estimate the printable codes for high quality printers even from the relatively small training datasets. Based on the performed experiments and obtained results we can identify three main criteria for successful fake detection: \textbf{(a)} the printing quality, \textbf{(b)} the amount of errors in estimated codes and \textbf{(c)} the regularity of the estimated errors. The defenders should prefer average quality printers with a dot-gain sufficient to make regular errors in the originals estimation. Moreover, the results show that modern machine learning technologies make the printable graphical codes vulnerable to clonability attacks.

For future work, we aim at examining other types of graphical codes, at investigating the possibilities of mobile phones for the detection of fake codes and to compare the abilities of machine learning approaches versus hand-crafted attacks. 
Finally, we plan to consider \textit{GAN}-like architectures to produce even more accurate fakes. The impact of the number of training examples and training from the original digital templates are also amongst our future priorities. 

\clearpage
\newpage
\bibliographystyle{IEEEbib}
\bibliography{refs}

\end{document}